\documentclass[prd,preprint,superscriptaddress]{revtex4}
\usepackage{amssymb}
\usepackage{revsymb}
\usepackage{graphicx}

\begin{document}
\title{Mini Black Holes and the Relic Gravitational Waves Spectrum}
\author{Germ\'{a}n Izquierdo \footnote{E-mail address: german.izquierdo@uab.es}}
\affiliation{Departamento de F\'{\i}sica, Universidad Aut\'{o}noma de Barcelona,
08193 Bellaterra (Barcelona) Spain}
\author{Diego Pav\'{o}n\footnote{E-mail address: diego.pavon@uab.es}}
\affiliation{Departamento de F\'{\i}sica, Universidad Aut\'{o}noma de Barcelona,
08193 Bellaterra (Barcelona) Spain}
\begin{abstract}
In this paper we explore the impact of an era -right after reheating- 
dominated by mini black holes and radiation on the spectrum of relic 
gravitational waves. This era may lower the spectrum several orders 
of magnitude.
\end{abstract}
\maketitle
\section{Introduction}
Gravitational wave detection is a very topical subject and great efforts are
underway in that direction. There are five cryogenic resonant-bar detectors
of gravitational waves in operation, sensitive ground-based
laser-interferometers will be operational soon (LIGO, GEO600 and VIRGO) \cite%
{ligo, geo, virgo} meanwhile TAMA300 is collecting data at their level of
sensitivity and the space-based Laser Interferometer Space Antenna (LISA) is
planned to be launched around the year 2011 \cite{Lisa}. Plans for advanced
ground-based interferometers are also being developed.

The most serious hurdle in the detection arises from the very weak
interaction of gravitational waves with matter and other fields. Only at
energies of order of the Planck scale ($\sim 10^{19}GeV$) will gravitational
interaction become as strong as the electromagnetic one. Paradoxically this
fact makes their detection\ so attractive, as it would give us information
about the epoch at which gravitational waves were decoupled from the
cosmological dynamics, i.e., the Planck time.

Gravitational waves can be produced by local sources, such as coalescing
stellar-mass black holes, compact binary stars and supernovae explosions, or
can have a cosmic origin, as the decay of cosmic strings and the
amplification of zero-point fluctuations due to the expansion of the
Universe. Waves of the latter type are usually called relic gravitational
waves (RGWs) and should form an isotropic stochastic background somehow
similar to the CMBR, but without a thermal distribution \cite{Allen96}. As
is well known, RGWs are a unavoidable consequence of general relativity and
quantum field theory in curved space-time.\ 

From the point of view of cosmologists obtaining data of RGWs spectrum will
be extremely interesting as it would make possible to reconstruct the scale
factor of the Universe. Different cosmic stages of expansion with different
equations of estate for the matter content will produce a well determined
power spectrum of RGWs. Even without data of the spectrum of RGWs, we can
still get useful information by analyzing their influence in well-known
processes as primordial nucleosynthesis and the anisotropy they would induce
in the CMBR. It is possible to test the validity of cosmological models (or
to restrict them) by comparing the theoretical spectrum of RGWs produced\
with the maximum values allowed by the CMBR anisotropy data and primordial
nucleosynthesis.

The RGWs production has been usually linked to inflation, though inflation
is not the only process leading to a RGWs spectrum \cite{Allen96, Buon}. By
contrast the recently proposed scenario of \ Khoury \textit{et al}. \cite%
{Khoury} does not lead to a spectrum of RGWs. The spectrum in an expanding
universe with a de Sitter inflationary era followed by a radiation dominated
era and finally by a dust dominated era was calculated by several authors 
\cite{grish93, Allen, Maia93} and a general expression for the creation of
RGWs in a multistage model was derived by Maia \cite{Maia93}. Other
cosmological models can be used to obtain in each case a spectrum of RGWs,
e.g. an initial inflationary stage different from the de Sitter one \cite%
{Turner}, a model with an intermediate additional era of stiff matter
between a power law inflation and the radiation era \cite{grish01}.

The aim of this paper is to calculate the power spectrum of RGWs in a
spatially flat Friedmann-Robertson-Walker universe that begins with de Sitter
era, followed by an era dominated by a mixture of mini black holes (MBHs)
and radiation, then a radiation dominated era (after the MBHs evaporated)
and finally a dust dominated era. We will also see what constraints on the
free parameters of the cosmological model can be drawn from the CMBR.

As it turns out, the era dominated by a mixture of MBHs and radiation leads
to a significantly reduced power spectrum of RGWs \ at present time. This
may have implications in case LISA\ does not detect a spectrum.

Except occasionally, we will use units for which $\hbar =c=k_{B}=1$.

\section{RGW spectrum in an expanding Universe}

\subsection{Basics of RGWs amplification}

\qquad We consider a flat FRW universe with line element 
\[
ds^{2}=-dt^{2}+a(t)^{2}\left[ dr^{2}+r^{2}d\Omega ^{2}\right] =a(\eta
)^{2}[-d\eta ^{2}+dr^{2}+r^{2}d\Omega ^{2}], 
\]%
where $t$ and $\eta $ are, respectively, the cosmic and conformal time ($%
a(\eta )d\eta ={dt}$).

\qquad By slightly perturbing the metric ($\overline{g}_{ij}=g_{ij}+h_{ij}$, 
$\left\vert h_{ij}\right\vert \ll \left\vert g_{ij}\right\vert $, $%
i,j=0,1,2,3$) the perturbed Einstein equations follow. To first order, the
transverse-traceless tensor solution which represents sourceless weak GWs
can be expressed as \cite{Lifs, grish74}%
\[
h_{\alpha \beta }(\eta ,{x})=\int h_{\alpha \beta }^{(k)}(\eta ,\mathbf{x}%
)d^{3}k, 
\]%
\begin{equation}
h_{\alpha \beta }^{(k)}(\eta ,\mathbf{x})=\frac{\mu (\eta )}{a(\eta )}%
G_{\alpha \beta }(\mathbf{k},\mathbf{x}),
\end{equation}%
where Greek indices run from $1$ to $3$, and $\mathbf{k}$ is the commoving
wave vector. The functions $G_{\alpha \beta }(\mathbf{k},\mathbf{x})$ and $%
\mu (\eta )$ satisfy the equations 
\begin{equation}
{G_{\alpha }^{\beta }}_{;\gamma }^{;\gamma }=-k^{2}G_{\alpha }^{\beta
},\qquad {G_{\alpha }^{\beta }}_{;\beta }=G_{\alpha }^{\alpha }=0,  \label{G}
\end{equation}%
\begin{equation}
\mu ^{\prime \prime }(\eta )+\left( k^{2}-\frac{a^{\prime \prime }(\eta )}{%
a(\eta )}\right) \mu (\eta )=0,  \label{eqmu}
\end{equation}%
where the prime indicates derivative with respect conformal time and $%
k=\left\vert \mathbf{k}\right\vert $ is the constant wave number related to the 
physical wavelength and frequency by $k=2\pi a/\lambda =2\pi af=a\ \omega .$

The functions ${G_{\alpha }^{\beta }}$ are combinations of $\exp (\pm i%
\mathbf{kx})$ which contain the two possible polarizations of the wave,
compatibles with the conditions (\ref{G}).

The equation (\ref{eqmu}) can be interpreted as an oscillator parametrically
excited by the term $a^{\prime \prime }/a$. When $k^{2}\gg \frac{a^{\prime
\prime }}{a}$, i.e., for high frequency waves, expression (\ref{eqmu})
becomes the equation of a harmonic oscillator whose solution is a free wave.
The amplitude of $h_{\alpha \beta }^{(k)}(\eta ,\mathbf{x})$ will decrease
adiabatically as $a^{-1}$ in an expanding universe. In the opposite regime,
when $k^{2}\ll \frac{a^{\prime \prime }}{a}$, the solution to (\ref{eqmu})
is a lineal combination of $\mu _{1}\propto a(\eta )$ {and }$\mu _{2}\propto
a(\eta )\int d\eta \ a^{-2}$. In an expanding universe $\mu _{1}$ grows
faster than $\mu _{2}$ and will soon dominate. Accordingly, the amplitude
of $h_{\alpha \beta }^{(k)}(\eta ,\mathbf{x})$ will remain constant so long
as the condition $k^{2}\ll \frac{a^{\prime \prime }}{a}$ is satisfied. When
it is no longer satisfied, the wave will have an amplitude greater than it
would in the adiabatic behavior. This phenomenon is known as \textquotedblleft
superadiabatic\textquotedblright\ or \textquotedblleft parametric
amplification\textquotedblright\ of gravitational waves \cite{grish93,
grish74}.

For power law expansion $a\propto \eta ^{l}$ ($l=-1,1,2$ for inflationary,
radiation dominated and dust dominated universes, respectively) the general
solution to equation (\ref{eqmu}) is 
\[
\mu (\eta )=(k\eta )^{\frac{1}{2}}\left( K_{1}J_{l-\frac{1}{2}}(k\eta
)+K_{2}J_{-\left( l-\frac{1}{2}\right) }(k\eta )\right) , 
\]%
where $J_{l-\frac{1}{2}}(k\eta )$ , $J_{-\left( l-\frac{1}{2}\right) }(k\eta
)$ are Bessel functions of the first kind and $K_{1,2}$ are integration
constants.

\subsection{Spectrum}

Here we succinctly recall the work of Grishchuk \cite{grish93} to obtain the
RGW spectrum in a universe that begins with a de Sitter stage, followed by a
radiation dominated era and a dust era till the present time. Transitions
between successive eras are assumed instantaneous.\ The scale factor is%
\begin{equation}
a(\eta )=\left\{ 
\begin{array}{c}
-\frac{1}{H_{1}\eta }\qquad (-\infty <\eta <\eta _{1}<0), \\ 
\frac{1}{H_{1}\eta _{1}^{2}}(\eta -2\eta _{1})\qquad (\eta _{1}<\eta <\eta
_{2}), \\ 
\frac{1}{4H_{1}\eta _{1}^{2}}\frac{(\eta +\eta _{2}-4\eta _{1})^{2}}{\eta
_{2}-2\eta _{1}}{\qquad }(\eta _{2}<\eta <\eta _{0})%
\end{array}%
\right.  \label{sclfac1}
\end{equation}%
where the subindexes $1,2$ correspond to the sudden transitions from
inflation to radiation era and \ from radiation to dust era, $H_{1}$
represents the Hubble factor at the end of the inflationary era and the
subindex zero indicates the present time.

The exact solution to the equation (\ref{eqmu}) for each era
specifies to
\begin{eqnarray}
\mu _{I}(\eta ) &=&C_{I}\left[ \cos (k\eta +\phi _{I})-\frac{1}{k\eta }\sin
(k\eta +\phi _{I})\right] \qquad \text{(inflationary era)}  \label{muinfl} \\
\mu _{R}(\eta ) &=&C_{R}\sin (k\eta _{R}+\phi _{r})\qquad \qquad \qquad
\qquad \qquad \qquad \text{(radiation era)} \\
\mu _{D}(\eta ) &=&C_{D}\left[ \cos (k\eta _{D}+\phi _{D})-\frac{1}{k\eta
_{D}}\sin (k\eta _{D}+\phi _{D})\right] \qquad \text{(dust era)},
\label{mudust}
\end{eqnarray}
where $C_{I,R,D}$, $\phi _{I,R,D}$ are constants of integration, $\eta
_{R}=\eta -2\eta _{1}$ and $\eta _{D}=\eta +\eta _{2}-4\eta _{1}$ .

It is possible to express $C_{R}$, $\phi _{R}$ \ and $C_{D}$, $\phi _{D}$ in
terms of $C_{I}$, $\phi _{I}$ and $C_{R},$ $\phi _{R}$ respectively as $\mu
(\eta )$ must be continuous at the transition times $\eta =\eta _{1}$ and $%
\eta =\eta _{2}$. Averaging the solution over the initial phase $\phi _{I}$
the amplification factor is found to be

\[
R(k)=\frac{C_{M}}{C_{I}}\sim \left\{ 
\begin{array}{c}
1\qquad (k\gg -1/\eta _{1}), \\ 
k^{-2}\qquad (-1/\eta _{1}\gg k\gg 1/(\eta _{D2})), \\ 
k^{-3}\qquad (1/(\eta _{D2})\gg k).%
\end{array}%
\right. 
\]

It is usual to consider that the initial vacuum spectrum of RGWs $h_{i}(k)$
is proportional to $k$ \cite{grish93}. The rationale behind that is the 
following. One assimilates the 
quantum zero-point fluctuations of vacuum with classical waves of certain 
amplitudes and arbitrary phases; consequently it is permissible to equalize  
$\hbar \omega/2$ with the energy density of the gravitational waves,
$c^{4}h^{2}/(G \lambda^{2})$ times $\lambda^{3}$. Thereby  
the initial vacuum spectrum of RGWs is given  by $h(k) \propto k$
\cite{grish93}, \cite{anny}.

Accordingly, the spectrum of RGWs in the dust era will be 
\begin{equation}
h_{f}(k)=R(k)h_{i}(k)\sim \left\{ 
\begin{array}{c}
k\qquad (k\gg -1/\eta _{1}), \\ 
k^{-1}\qquad (-1/\eta _{1}\gg k\gg 1/(\eta _{D2})), \\ 
k^{-2}\qquad (1/(\eta _{D2})\gg k).%
\end{array}%
\right.  \label{speclas}
\end{equation}

\subsection{ Bogolubov coefficients}

The classical amplification mechanism may be seen as a spontaneous particle
creation by the gravitational field acting on the initial quantum vacuum 
\cite{Allen, Maia93, Birrell, Parker}. This is possible as the GW equation
is not conformally invariant when $a^{\prime \prime }/a\neq 0$ even in an
isotropic background \cite{grish74}.

The GW equation may be interpreted as the massless Klein-Gordon equation 
\cite{Allen}. Its solution can be written as%
\begin{equation}
h_{ij}(\eta ,\mathbf{x})=\int \left( A_{(k)}h_{ij}^{(k)}(\mathbf{k}%
,x)+A_{(k)}^{\dag }h_{ij}^{(k)}(\mathbf{k},x)\right) d^{3}k,  \label{fieldh}
\end{equation}%
\[
h_{ij}^{(k)}(\mathbf{k},x)=\frac{1}{\sqrt{\pi }}e_{ij}(\mathbf{k})\frac{\mu
_{(k)}(\eta )}{a(\eta )}e^{i\mathbf{k\cdot x}}\mathbf{,} 
\]%
where $A_{(k)}$, $A_{(k)}^{\dag }$ are the annihilation and creation
operators, respectively, $e_{ij}(\mathbf{k})$ contains the two possible
polarizations of the wave and $\mu _{(k)}(\eta )$ is a solution to the
equation (\ref{eqmu}), as in the classical approach, but with the additional
condition%
\begin{equation}
\mu _{(k)}\mu _{(k)}^{\ast \prime }-\mu _{(k)}^{\ast }\mu _{(k)}^{\prime }=i,
\label{condcuan}
\end{equation}%
which comes from the commutation relations of the operators $A_{(k)}$, $%
A_{(k)}^{\dag }$ and the definition of the field $h_{ij}(\eta ,\mathbf{x})$ 
\cite{Parker, Birrell}.

The solution (\ref{fieldh}) can be expressed in terms of another family of
orthogonal modes $\bar{\mu}_{(k)}(\eta )$ as, in contrast with the
Mikowskian space-time, in a curved space-time there is no privileged family.
The two families of modes are related by a Bogolubov transformation%
\[
\mu _{(k)}(\eta )=\alpha \bar{\mu}_{(k)}(\eta )+\beta \bar{\mu}_{(k)}^{\ast
}(\eta ), 
\]%
where $\alpha $, $\beta $ are the Bogolubov coefficients. The number of
particles of the family $\mu _{(k)}$, $N_{k}$, contained in the vacuum state
of the modes $\bar{\mu}_{(k)}$ is given by 
\[
\langle \overline{0}|N_{k}|\overline{0}\rangle =\langle \overline{0}%
|A_{(k)}^{\dagger }A_{(k)}|\overline{0}\rangle =|\beta |^{2}. 
\]

It is necessary to ascertain which of the solutions to (\ref{eqmu})
correspond to real particles. An approach to that problem is known as the
adiabatic vacuum approximation \cite{Birrell}. Basically, it assumes that the
curved space-time is asymptotically Mikowskian in the limit $k\rightarrow
\infty $. In that limit, the creation-destruction operators of each family
exactly correspond  to those associated to real particles and 
consequently $\alpha =1$ and $\beta =0$ as the two families of modes 
represent the same vacuum state in the Mikowskian space-time. 
This argument is identical to that made in section II.A, solutions to
equation (\ref{eqmu}) with $k\gg a^{\prime \prime }/a$ are free waves
and they do not experience amplification.

Following Allen \cite{Allen} and Maia \cite{Maia93}, we will now evaluate
the number of RGWs created from the initial vacuum state in an expanding
universe. In the three-stage cosmological model we are considering with the
scale factor given by (\ref{sclfac1}), the initial state is the vacuum
associated with the modes of the inflationary stage $\mu _{I}(\eta )$, which
are a solution to the equation (\ref{eqmu}) compatible with the condition (%
\ref{condcuan}). Taking into account the shape of the scale factor at this
era the modes are 
\begin{equation}
\mu _{I}=(\sqrt{\pi }/2)e^{i\psi _{I}}k^{-1/2}x^{1/2}H_{-3/2}^{(2)}(x),
\label{muinfcuant}
\end{equation}%
where $x=k\eta $ and $\psi _{I}$ is an arbitrary constant phase and $%
H_{-3/2}^{(2)}(x)$ is the Hankel function of order $-3/2$. The proper modes
of the radiation era are%
\begin{equation}
\mu _{R}=(\sqrt{\pi }/2)e^{i\psi
_{R}}k^{-1/2}x_{R}^{1/2}H_{1/2}^{(2)}(x_{R}),  \label{murad}
\end{equation}%
where $x_{R}=k(\eta -2\eta _{1})$ and $\psi _{R}$ is again a constant phase.

The two families of modes are related by%
\begin{equation}
\mu _{I}(\eta )=\alpha _{1}\mu _{R}(\eta )+\beta _{1}\mu _{R}^{\ast }(\eta ).
\label{bogtr}
\end{equation}%
From the continuity of $\mu (\eta )$ at the transition time $\eta _{1}$ we
obtain 
\begin{equation}
\alpha _{1}=-1+\frac{i}{k\eta _{1}}+\frac{1}{2(k\eta _{1})^{2}},\qquad \beta
_{1}=\frac{1}{2(k\eta _{1})^{2}},  \label{coef1}
\end{equation}%
where we have neglected an irrelevant phase. The modes with frequency at the
transition time $f=2\pi k/a(\eta _{1})$ larger than the characteristic time
scale of transition are exponentially suppressed. The characteristic time
scale is usually taken to be the inverse of the Hubble factor at the
transition, $H_{1}^{-1}$ in this case. The coefficients will be $\alpha
_{1}=1$ and $\beta _{1}=0$ for RGWs with $k>2\pi a_{1}H_{1}$ and (\ref{coef1}%
) when $k<2\pi a_{1}H$.

In the dust era ($\eta >\eta _{2}$) the solution for the modes is%
\begin{equation}
\mu _{D}=(\sqrt{\pi }/2)e^{i\psi
_{D}}k^{-1/2}x_{D}^{1/2}H_{3/2}^{(2)}(x_{D}),  \label{mudustcuan}
\end{equation}%
where $x_{D}=k\left( \eta +\eta _{2}-4\eta _{1}\right) $ and is related to
the radiation ones by%
\begin{equation}
\mu _{R}(\eta )=\alpha _{2}\mu _{D}(\eta )+\beta _{2}\mu _{D}^{\ast }(\eta ).
\label{bogtrr_d}
\end{equation}%
Similarly one obtains%
\[
\alpha _{2}=-i\left( 1+\frac{i}{2k(\eta _{2}-2\eta _{1})}-\frac{1}{8\left(
k(\eta _{2}-2\eta _{1})\right) ^{2}}\right) ,\qquad \beta _{2}=\frac{i}{%
8\left( k(\eta _{2}-2\eta _{1})\right) ^{2}}, 
\]%
for $k<2\pi a(\eta _{2})H_{2}$ and $\alpha _{2}=1$, $\beta _{2}=0$ for $%
k>2\pi a(\eta _{2})H_{2}$ where $H_{2}$ is the Hubble factor evaluated at
the transition $\eta _{2}$.

In order to relate the modes of the inflationary era to the modes of the
dust era, we make use of the total Bogolubov coefficients $\alpha _{Tr2}$
and $\beta _{Tr2}$ \cite{Allen, Maia93}. For $k>2\pi a_{1}H_{1}$, we find
that $\alpha _{Tr2}=1$, $\beta _{Tr2}=0$; in the range $2\pi
a_{1}H_{1}>k>2\pi a_{2}H_{2}$, the coefficients are $\alpha _{Tr2}=\alpha
_{1}$ and $\beta _{Tr2}=\beta _{1}$, and finally for $k<2\pi a(\eta _{2})H_{2}$
we obtain that%
\[
\beta _{Tr2}=-\frac{2(\eta _{2}-2\eta _{1})+\eta _{1}}{8k^{3}\eta
_{1}^{2}(\eta _{2}-2\eta _{1})^{2}}. 
\]

Thus the number of RGWs at the present time $\eta _{0}$ created from the
initial vacuum state is $\left\langle N_{\omega }\right\rangle =\left\vert
\beta _{Tr2}\right\vert ^{2}\sim \omega ^{-6}(\eta _{0})$ for $\omega (\eta
_{0})<2\pi (a_{2}/a_{0})H_{2}$, $\omega ^{-4}(\eta _{0})$ for $2\pi
(a_{1}/a_{0})H_{1}>\omega (\eta _{0})>2\pi (a_{2}/a_{0})H_{2}$ and zero for $%
\omega (\eta _{0})>2\pi (a_{1}/a_{0})H_{1}$, where we have used the present
value of the frequency, $\omega (\eta _{0})=k/a_{0}$.

Assuming that each RGW has an energy $2\hbar \omega (\eta )$, it is possible
to express the energy density of RGWs with frequencies in the range $\left[
\omega (\eta ),\omega (\eta )+d\omega (\eta )\right] $ as 
\begin{equation}
d\rho _{g}(\eta )=2\hbar \omega (\eta )\left[ \frac{\omega ^{2}(\eta )}{2\pi
^{2}c^{3}}d\omega (\eta )\right] \left\langle N_{\omega }\right\rangle
=P(\omega (\eta ))d\omega (\eta ),  \label{densquan}
\end{equation}%
where $P(\omega (\eta ))=\left( \omega ^{3}(\eta )/\pi ^{2}\right)
\left\langle N_{\omega }\right\rangle $ denotes the power spectrum. As the
energy density is a locally defined quantity, $\rho _{g}$ loses its meaning
for metric perturbations with wave length $\lambda =2\pi /\omega (\eta )$
larger than the Hubble radius $H^{-1}(\eta )$. The present power spectrum of
RGWs predicted by this model is
\begin{equation}
P\left( \omega \right) \sim \left\{ 
\begin{array}{c}
0\text{\qquad }\left( \omega (\eta _{0})>2\pi (a_{1}/a_{0})H_{1}\right) , \\ 
\omega ^{-1}(\eta _{0})\text{\qquad }\left( 2\pi (a_{2}/a_{0})H_{2}<\omega
(\eta _{0})<2\pi (a_{1}/a_{0})H_{1}\right) , \\ 
\omega ^{-3}(\eta _{0})\text{\qquad }\left( 2\pi H_{0}<\omega (\eta
_{0})<2\pi (a_{2}/a_{0})H_{2}\right) .%
\end{array}%
\right.  \label{espectr}
\end{equation}%
We can compare these results with the classical ones of the previous
subsection. The energy density of created RGWs in the classical approach is
defined from the spectrum as \cite{anny} 
\begin{equation}
\rho _{g}\sim \int_{2\pi a_{0}H_{0}}^{-1/\eta _{1}}h_{f}^{2}(k)kdk.
\label{densclas}
\end{equation}%
From equations (\ref{densquan}) and (\ref{densclas}) it follows 
\[
P(k)\sim kh_{f}^{2}(k). 
\]%
Thus both descriptions agree with regard to the dependence on $h_{f}$ and
the ranges of the wave number of the spectrum of RGWs, as for a power law\
scale factor the Hubble factor is $H(\eta )=l(\eta a(\eta ))^{-1}$.

\section{RGWs in a FRW universe with an era of\ Mini Black Holes and
radiation}

In this section we evaluate the spectrum of RGWs using the method of the
Bogolubov coefficients in a more general model than the conventional
three-stage model of the previous section.

As is well known, mini black holes can be created by quantum tunneling from
the hot radiation \cite{gross}; some cosmological consequences of this effect
have been studied \cite{Hayward, Barrow}. We shall assume these mini black
holes (MBHs) are created right after the inflationary period (once the
reheating is accomplished) and coexist with the radiation until they
evaporate. During that era the total energy density can be
approximated by $\rho =\rho _{BH}+\rho _{R}$ and the total pressure is 
\begin{equation}
p=p_{BH}+p_{R}=(\gamma -1)\rho ,  \label{eqstate}
\end{equation}%
where $1\leq \gamma <4/3$. MBHs follows the dust equation of state $p_{BH}=0$
as they can be considered non-relativistic matter. If the density of MBHs is
large enough to dominate the expansion of the Universe, then $\gamma \simeq
1 $. In the opposite case, the Universe expansion is dominated by the
radiation, $\gamma \simeq 4/3$ and the Universe undergoes the three stages
of the previous section. From the Einstein equations and (\ref{eqstate}) one
finds $a(\eta )\propto \eta ^{l}$, where $l = 27(3\gamma -2) 
(\Rightarrow 1<$ $l\leq 2)$ during the
\textquotedblleft MBHs+rad\textquotedblright\ era. The MBHs eventually
evaporate in relativistic particles after some time span whose duration is
model dependent.

\subsection{Power Spectrum}

Assuming that the expansion of the Universe is ab initio dominated by the
vacuum energy of some scalar field (the inflaton), then dominated by the
mixture of MBHs\ and radiation, later radiation dominated (after the
evaporation of the MBHs), and finally dust dominated up to the present time,
the scale factor of each era is%
\[
a(\eta )=\left\{ 
\begin{array}{c}
-\frac{1}{H_{1}\eta }\qquad (-\infty <\eta <\eta _{1}<0),\text{\qquad
(inflation)} \\ 
\frac{\left[ \eta _{BH}\right] ^{l}}{l^{l}H_{1}(-\eta _{1})^{l+1}}\qquad
(\eta _{1}<\eta <\eta _{2}),\text{\qquad (\textquotedblleft
MBHs+rad\textquotedblright\ era)} \\ 
\frac{\left( \eta _{R2}\right) ^{l-1}}{H_{1}(-\eta _{1})^{l+1}}\eta _{R}{%
\qquad }(\eta _{2}<\eta <\eta _{3}),\text{\qquad (radiation\ era)} \\ 
\frac{\left( \eta _{R2}\right) ^{l-1}}{2H_{1}(-\eta _{1})^{l+1}\eta _{D3}}%
\left[ \eta _{D}\right] ^{2}{\qquad }(\eta _{3}<\eta <\eta _{0}),\text{%
\qquad (dust era)}%
\end{array}%
\right. 
\]%
where $\eta _{BH}=\eta -(l+1)\eta _{1}$, $\eta _{R}=\eta +\frac{(1-l)}{l}%
\eta _{2}-\frac{(l+1)}{l}\eta _{1}$, $\eta _{D}=\eta +\eta _{3}+2\frac{(1-l)%
}{l}\eta _{2}-2\frac{(l+1)}{l}\eta _{1}$, $\eta _{R2}=\left[ \eta
_{2}-(l+1)\eta _{1}\right] /l$ and $\eta _{D3}=2\left[ \eta _{3}+\frac{(1-l)%
}{l}\eta _{2}-\frac{(l+1)}{l}\eta _{1}\right] $. As in the previous section,
the transitions betwen stages are assumed to be instantaneous.

The shape of $\mu (\eta )$ can be found by solving the equation (\ref{eqmu})
in each era. For the de Sitter era $\mu (\eta )$ is given by (\ref{muinfcuant}) 
as above. For the \textquotedblleft MBHs+rad \textquotedblright\ era the solution 
of (\ref{eqmu}) is

\[
\mu _{BH}=(\sqrt{\pi }/2)e^{i\psi
_{BH}}k^{-1/2}x_{BH}^{1/2}H_{l-1/2}^{(2)}(x_{BH}), 
\]%
where $x_{BH}=k\,\eta _{BH},$ and it is related to the modes of inflation by
the Bogolubov transformation (\ref{bogtr}) with $\mu _{R}$ replaced by $\mu
_{BH}$. By evaluating the Bogolubov coefficients at the transition we obtain 
\[
\alpha _{1}^{l},\text{\ }\beta _{1}^{l}\simeq \frac{l^{2}2^{l}}{\left(
-l\eta _{1}\right) ^{l+1}}k^{-(l+1)} 
\]%
when $k<2\pi a_{1}H_{1}$ and $\alpha _{1}^{l}=1,$ $\beta _{1}^{l}=0$ when $%
k>2\pi a_{1}H_{1}$.

The solution for the radiation era is again (\ref{murad}) with $%
x_{R}=k\,\eta _{R}$. The coefficients that relate (\ref{murad}) with $\mu
_{BH}$ are 
\[
\alpha _{2}^{l}=-\frac{1}{2}\sqrt{\frac{\pi lx_{R2}}{2}}\left[ \left( \frac{1%
}{x_{R2}}-i\right) H_{l-\frac{1}{2}}^{(2)}(lx_{R2})-H_{l+\frac{1}{2}%
}^{(2)}(lx_{R2})\right] e^{ix_{R2}}, 
\]%
\[
\beta _{2}^{l}=\frac{1}{2}\sqrt{\frac{\pi lx_{R2}}{2}}\left[ \left( \frac{1}{%
x_{R2}}+i\right) H_{l-\frac{1}{2}}^{(2)}(lx_{R2})-H_{l+\frac{1}{2}%
}^{(2)}(lx_{R2})\right] e^{-ix_{R2}}, 
\]%
when $k<2\pi a_{2}H_{2}$ and $\alpha _{2}^{l}=1,$ $\beta _{2}^{l}=0$ when $%
k>2\pi a_{2}H_{2}.$

The modes of the dust era are given by (\ref{mudustcuan}) with $%
x_{D}=k\,\eta _{D}$ and are related to the modes of the radiation era by the
coefficients%
\[
\alpha _{3}=-i\left( 1+\frac{i}{x_{D3}}-\frac{1}{2x_{D3}^{2}}\right) ,\qquad
\beta _{3}=i\frac{1}{2x_{D3}^{2}}, 
\]%
when $k<2\pi a_{3}H_{3}$ and $\alpha _{3}=1,$ $\beta _{3}=0$ when $k>2\pi
a_{3}H_{3}.$

In order to evaluate the spectrum of RGWs the total Bogolubov\ coefficients
are needed. The coefficients relating the initial vacuum state with the
modes of the radiation state can be evaluated with the help of the
relationships 
\[
\alpha _{Tr2}^{l}=\alpha _{2}^{l}\alpha _{1}^{l}+\beta _{2}^{l\ast }\beta
_{1}^{l},\qquad \beta _{Tr2}^{l}=\beta _{2}^{l}\alpha _{1}^{l}+\alpha
_{2}^{l\ast }\beta _{1}^{l}, 
\]%
and one obtains 
\begin{equation}
\alpha _{Tr2}^{l},\text{ }\beta _{Tr2}^{l}\simeq \left\{ 
\begin{array}{c}
1,\text{ }0\text{\qquad }\left( k>2\pi a_{1}H_{1}\right) , \\ 
\alpha _{1}^{l},\text{ }\beta _{1}^{l}\text{\qquad }\left( 2\pi
a_{1}H_{1}>k>2\pi a_{2}H_{2}\right) , \\ 
\frac{l^{-l+2}(2l^{2}-3l+1)}{8\left( -l\eta _{1}\right) ^{l+1}\left( \eta
_{R2}\right) ^{l-1}}k^{-2l}\text{\qquad }\left( k<2\pi a_{2}H_{2}\right) .%
\end{array}%
\right.  \label{btr2bh}
\end{equation}%
Finally, the total coefficients that relate the inflationary modes with the
modes of the dust era evaluated from 
\[
\alpha _{Tr3}^{l}=\alpha _{3}\alpha _{Tr2}^{l}+\beta _{3}^{\ast }\beta
_{Tr2}^{l},\qquad \beta _{Tr3}^{l}=\beta _{3}\alpha _{Tr2}^{l}+\alpha
_{3}^{\ast }\beta _{Tr2}^{l} 
\]%
are found to be 
\begin{equation}
\alpha _{Tr3}^{l},\text{ }\beta _{Tr3}^{l}\simeq \left\{ 
\begin{array}{c}
\alpha _{Tr2}^{l},\text{ }\beta _{Tr2}^{l}\text{\qquad }\left( k>2\pi
a_{3}H_{3}\right) , \\ 
\frac{l^{-l+2}(2l^{2}-3l+1)}{8\left( -l\eta _{1}\right) ^{l+1}\left( \eta
_{R2}\right) ^{l-1}\eta _{D3}}k^{-(2l+1)}\text{\qquad }\left( k<2\pi
a_{3}H_{3}\right) .%
\end{array}%
\right.  \label{btr3bh}
\end{equation}

We are now in position to calculate the current spectrum of RGWs. Taking
into account that 
\[
-\eta _{1}=(a_{1}H_{1})^{-1},\qquad \eta
_{R2}=(a_{2}/a_{1})^{1/l}(a_{1}H_{1})^{-1},\qquad \eta
_{D3}=2(a_{3}/a_{2})(a_{2}/a_{1})^{1/l}(a_{1}H_{1})^{-1}, 
\]%
and $\omega =k/a_{0}$, the power spectrum $P(\omega )$ can be written as%
\begin{equation}
\begin{array}{l}
P\simeq 0{\qquad }\left( \omega >2\pi \frac{a_{1}}{a_{0}}H_{1}\right) , \\ 
\\ 
P\simeq \frac{l^{2-2l}2^{2l}}{\pi ^{2}}\left( \frac{a_{1}}{a_{0}}\right)
^{2l+2}H_{1}^{2l+2}\omega ^{-(2l-1)}{\qquad }\left( 2\pi \frac{a_{1}}{a_{0}}%
H_{1}>\omega >2\pi \frac{a_{2}}{a_{0}}H_{2}\right) , \\ 
\\ 
P\simeq \frac{l^{2-4l}(2l^{2}-3l+1)^{2}}{64\pi ^{2}}\left( \frac{a_{1}}{a_{0}%
}\right) ^{4l}\left( \frac{a_{1}}{a_{2}}\right) ^{2-2/l}H_{1}^{4l}\omega
^{-(4l-3)}{\qquad }\left( 2\pi \frac{a_{2}}{a_{0}}H_{2}>\omega >2\pi \frac{%
a_{3}}{a_{0}}H_{3}\right) , \\ 
\\ 
P\simeq \frac{l^{2-4l}(2l^{2}-3l+1)^{2}}{16\pi ^{2}}\left( \frac{a_{1}}{a_{0}%
}\right) ^{4l+4}\left( \frac{a_{0}}{a_{3}}\right) ^{2}H_{1}^{4l+2}\omega
^{-(4l-1)}{\qquad }\left( 2\pi \frac{a_{3}}{a_{0}}H_{3}>\omega >2\pi
H_{0}\right) .%
\end{array}
\label{espectrbh}
\end{equation}

Comparing the power of $\omega $ in (\ref{espectr}) and (\ref{espectrbh})
for $\omega <2\pi \frac{a_{1}}{a_{0}}H_{1}$, we conclude that the four-stage
scenario leads to a higher number of RGWs created at low frequencies than the
three-stage scenario. This fact can be explained intuitively with the
classical amplification approach. In the three-stage scenario the RGWs are parametrically 
amplified as long as $k^{2}<a^{\prime \prime }(\eta)/a(\eta )$. For $\eta=\eta _{1}$, 
$a^{\prime \prime }(\eta )/a(\eta )$ vanishes and there is no more amplification. 
On the other hand, in the four-stage scenario the RGWs with $\omega <2\pi \frac{a_{1}}{a_{0}}H_{1}$
are amplified until $\eta _{1}$ by the same term $a^{\prime \prime }(\eta
)/a(\eta )=2/\eta^2$ than in the three stage model 
and from $\eta_{1}$ to $\eta_{2}$ by the term $l(l-1)/\eta_{BH}^2$. Consequently 
they have a larger amplitude in the radiation era.  

\begin{figure}[tbp]
\includegraphics*[angle=-90,scale=0.75]{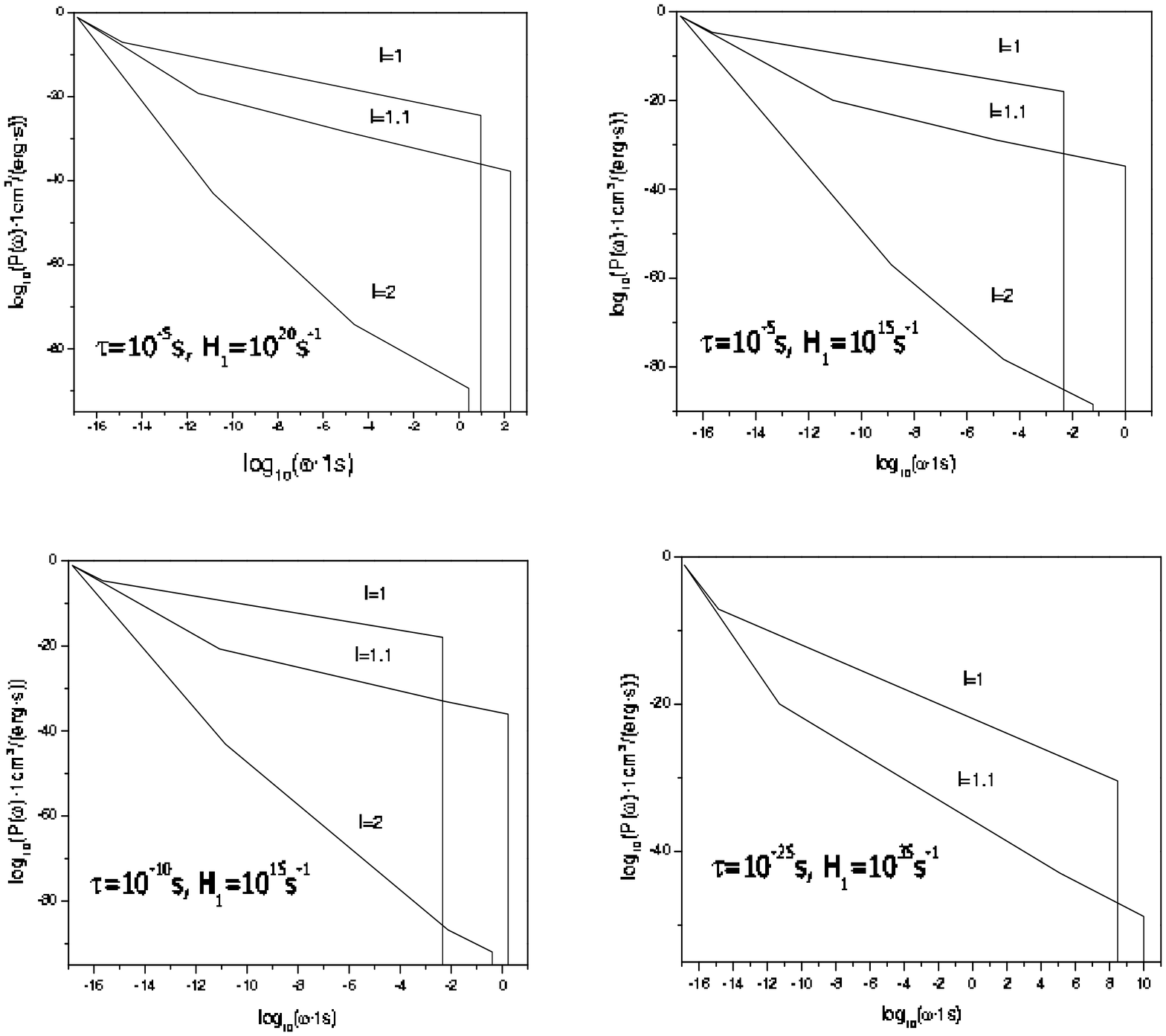}
\caption{Spectrum of RGWs in an expanding universe with a `MBHs+rad' era for
certain values of $l$, $\protect\tau $ and $H_{1}$. The spectrum predicted
for the three-stage model of the previous section is plotted for comparison, 
$l=1$. It is assumed that each spectrum has the maximum value allowed by the
CMBR anisotropy data at the frecuency $\omega=2\pi H_0=2.24\times 10^{-18}s^{-1}$. 
In the bottom-right panel the power spectrum with $l=2$ is excluded as it yields a CMBR
anisotropy larger than the observed.}
\label{figspec}
\end{figure}

Figure \ref{figspec} shows the spectrum (\ref{espectrbh}) for $l=2$ and $l=1.1$. 
As is apparent the four-stage scenario gives rise to a much lower
power spectrum than the three-stage scenario assuming that in each case the
spectrum has the maximum value allowed by the CMBR bound. The higher the MBHs
contribution to the energy density, the lower the final power spectrum.

In this four-stage cosmological scenario, the Hubble factor $H(\eta )$
decreases monotonically, while the energy density of the RGWs for $\eta
>\eta _{3}$ can be approximated by $\rho _{g}(\eta )\sim H^{-4l+2}(\eta )$
thereby it increases with expansion \cite{Maia93}. Obviously this scenario
will break down before $\rho _{g}(\eta )$ becomes comparable to the energy
density of matter and/or radiation since from that moment on the linear
approximation on which our approach is based ceases to be valid. The
parameters of our model will be constrained such that this cannot happen
before the current era.

\subsection{Evaluation of the parameters}

At this stage it is expedient to evaluate the parameters occurring in (\ref%
{espectrbh}). The redshift $\frac{a_{0}}{a_{3}}$, relating the present value
of the scale factor with the scale factor at the transition radiation-dust,
may be taken as $10^{4}$ \cite{Peebles}. The Hubble factor $H_{1}$ is
connected to the density at the inflationary era by 
\[
\rho _{1}=\frac{c}{\hbar }\frac{3m_{Pl}^{2}}{8\pi }H_{1}^{2}, 
\]%
where we have restored momentarily the fundamental constants. In any
reasonable model the energy density at that time must be larger than the
nuclear density ($\sim 10^{35}erg/cm^{3}$) and lower than the Planck density
($\sim 10^{115}erg/cm^{3}$) \cite{grish93b}, therefore

\begin{equation}
10^{3}s^{-1}<H_{1}<10^{43}s^{-1}.  \label{H1}
\end{equation}

Using the expression of the scale factor at the `MBHs+rad' era in terms of
the proper time, one obtains 
\begin{equation}
\frac{a_{2}}{a_{1}}=\left( 1+H_{1}\frac{l+1}{l}\tau \right) ^{l/(l+1)},
\label{a2/a1}
\end{equation}%
where $\tau $ is the time span of the `MBHs+rad' era which depends on the
evaporation history of the MBHs. The simplest model assumes that the MBHs
are all created instantaneously at some given time with the same mass and
terminate their evaporation simultaneously. If the MBHs evaporated freely,
their mass would evolve according to $\left( dM/dt\right) _{free}=-g_{\ast
}m_{Pl}^{4}/(3M^{2})$, \cite{Hawk}, where $g_{\ast }$ is the number of
particles for the black hole to evaporate into. However, it is natural to
assume that the MBHs are surrounded by an atmosphere of particles in quasi
thermal equilibrium with them. Therefore we have $\left| \left( dM/dt \right)_{atm} \right|\ll 
\left| \left( dM/dt\right) _{free} \right|$ for a comparatively long period whence the span
of the \textquotedblleft MBHs +rad\textquotedblright\ era is larger than if the MBHs evaporated freely. 
Other possible model consists in considering that MBHs are created according to
some nucleation rate during the era, instead of being all \ nucleated
instantaneously. These possibilities give us some freedom on evaluating $%
\tau $.

However, the `MBHs+rad' era span should be longer than the duration of the transition
at $\eta _{1}$ (as the transition is assumed instantaneous) in calculating
the spectrum of RGWs. To evaluate the adiabatic vacuum cutoff for the
frequency we have considered that the transition between whatever two
successive stages has a duration of the same order as the inverse of the
Hubble factor. This places the additional constraint 
\begin{equation}
\tau >H_{1}^{-1}.  \label{tau}
\end{equation}

Finally, $\frac{a_{1}}{a_{0}}$ can be evaluated from the evolution of the
Hubble factor until the present time 
\[
H_{0}=\left( \frac{a_{0}}{a_{3}}\right) ^{1/2}\left( \frac{a_{2}}{a_{1}}%
\right) ^{(l-1)/l}\left( \frac{a_{1}}{a_{0}}\right) ^{2}H_{1}. 
\]%
The current value of the Hubble factor is estimated to be $2.24\times
10^{-18}s^{-1}$ \cite{Spergel} and 
\begin{equation}
\left( \frac{a_{1}}{a_{0}}\right) ^{2}=\left( \frac{a_{3}}{a_{0}}\right)
^{1/2}\left( 1+H_{1}\frac{l+1}{l}\tau \right) ^{(1-l)/(l+1)}\frac{H_{0}}{%
H_{1}}.  \label{a1/a0}
\end{equation}

The only free parameters considered here are $l,$ $\tau $ and $H_{1}$, with
the restrictions $1<l\leq 2$, (\ref{tau}) and (\ref{H1}). The two first free
parameters depend on the assumption made on the MBHs, although it is
possible to obtain rigorous constraints on $H_{1}$ and $\tau $ from the
density of the RGWs.

\subsection{Restrictions on the \textquotedblleft
MBHs+rad\textquotedblright\ era}

It is obvious that $\rho _{g}$ cannot be arbitrarily large, in fact the RGWs
are seen as linear perturbations of the metric. The linear approximation
holds only for $\rho _{g}(\eta )\ll \rho (\eta )$, $\rho (\eta )$ being the
total energy density of the Universe. Several observational data place
constraints on $\rho _{g}$. The regularity of the pulses of stable
millisecond pulsars sets a constraint at frequencies of order $10^{-8}Hz$ 
\cite{Thors}. Likewise, there is a certain maximum value for $\rho _{g}$
compatible with the primordial nucleosynthesis scenario. But the most severe
constraints come from the high isotropy degree of the CMBR. We will focus on
the latter constraint. Metric perturbations with frequencies between $%
10^{-16}$ and $10^{-18}$ $Hz$ at the last scattering surface can produce
thermal fluctuations in the CMBR due to the Sachs-Wolfe effect \cite{sachs}.
These thermal fluctuations cannot exceed the observed value of $\delta
T/T\sim 5\times 10^{-6}$.

A detailed analysis of the CMBR bound yields \cite{Allen96, Allen Koranda}%
\begin{equation}
\Omega _{g}h_{100}^{2}<7\times 10^{-11}\left( \frac{H_{0}}{f}\right) ^{2}{%
\qquad ({H}_{0}<f<30\times H}_{0}{)}  \label{condPCMB}
\end{equation}%
where $\Omega _{g}=fP(f)/\rho _{0},$ $\rho _{0}=3cm_{Pl}^{2}H_{0}^{2}/(8\pi
\hbar )$ and $H_{0}=h_{100}^{{}}\times 100km/(s\times Mpc)$ and $%
h_{100}^{{}}=0.7$. The CMBR bound for the spectrum (\ref{espectrbh})
evaluated at the frequency $\omega =2\pi H_{0}$ reads \footnote{%
It is necessary to multiply (\ref{espectrbh}) for $\hbar /c^{3}$ in order to
obtain the correct units.} 
\[
1>\left( 2\pi \right) ^{-4l}l^{2-4l}(2l^{2}-3l+1)^{2}\left( \frac{a_{1}}{%
a_{0}}\right) ^{4l+4}\left( \frac{a_{0}}{a_{3}}\right) ^{2}\left( \frac{H_{1}%
}{3.72\times 10^{19}s^{-1}}\right) ^{3}\left( \frac{H_{1}}{H_{0}}\right)
^{4l-1}, 
\]%
and consequently
\begin{equation}
f(l,H_{1},\tau )=-107.69+l\left( 28.10+2\log _{10}\left( \frac{H_{1}}{1s^{-1}%
}\right) \right) -(2l-2)\log _{10}\left( 1+\frac{l+1}{l}H_{1}\tau \right)
\label{eqH1taul}
\end{equation}%
\[
+(-4l+2)\log _{10}(l)+2\log _{10}(2l^{2}-3l+1)<0. 
\]%
We next consider different values for $l$ and $\tau $.

\begin{figure}[tbp]
\includegraphics*[angle=-90,scale=0.7]{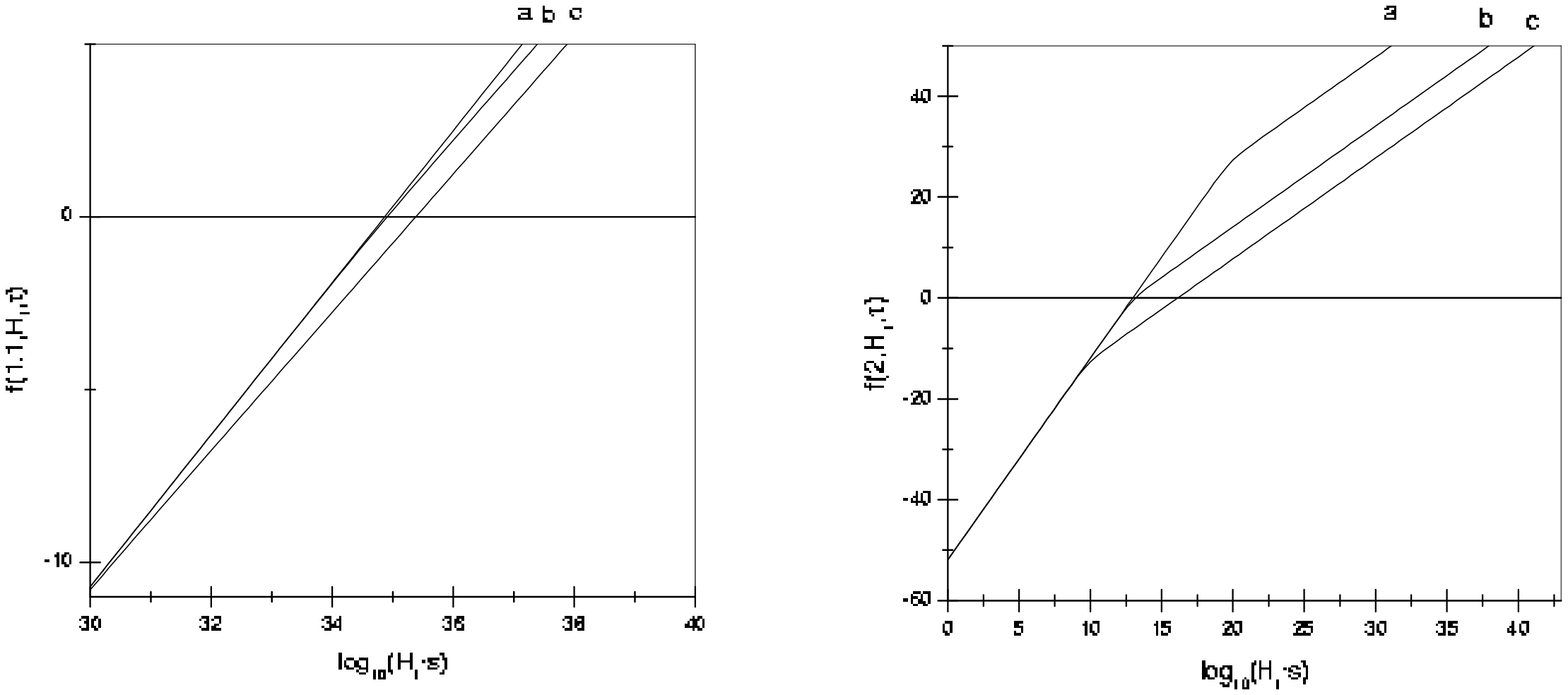}
\caption{{}The left panel depicts $f(1.1,H_{1},\protect\tau )vs.\log
_{10}H_{1}$ for a) $\protect\tau =10^{-40}s$, b) $\protect\tau =\protect\tau %
_{c}^{l=1.1}=1.23\times 10^{-35}s$, and c) $\protect\tau =10^{-30}s$. The
right panel depicts $f(2,H_{1},\protect\tau )$ for a) $\protect\tau %
=10^{-20}s$, b) $\protect\tau =\protect\tau _{c1}^{l=2}=6.75\times 10^{-14}s$%
, and c) $\protect\tau =10^{-10}s$. Conditions (\protect\ref{eqH1taul}) and
(\protect\ref{tau}) are satisfied for certain ranges of $\protect\tau $ and 
$H_{1}$ in each case. }
\label{graphf}
\end{figure}

(i) When $l=1.1$, the relation (\ref{eqH1taul}) reads%
\begin{equation}
f(1.1,H_{1},\tau )=-76.71+2.20\log _{10}\left( \frac{H_{1}}{1s^{-1}}\right)
-0.2\log _{10}\left( 1+\frac{l+1}{l}H_{1}\tau \right) <0,
\label{eqH1taul=1.1}
\end{equation}%
see Fig. \ref{graphf}. From it we observe that:

\begin{enumerate}
\item For $\tau <\tau _{c}^{l=1.1}=1.22\times 10^{-35}s$, the condition (\ref%
{eqH1taul=1.1}) is satisfied if $H_{1}<8.13\times 10^{34}s^{-1}$ and
conflicts with (\ref{tau}), which in the most favorable case is $%
H_{1}=8.13\times 10^{34}s^{-1}$ for $\tau =\tau _{c}^{l=1.1}$. For $l=1.1$,
there is no compatibility with the observed CMBR anisotropy when $\tau <\tau
_{c}^{l=1.1}$. Thus, this range of $\tau $ is ruled out.

\item For $\tau >\tau _{c}^{l=1.1}$, one obtains $H_{1}<H_{c}^{l=1.1}(\tau )$ from 
the condition (\ref{eqH1taul=1.1}). $H_{c}^{l=1.1}(\tau )$ is
always larger than $\tau ^{-1}$ in the range considered, e.g. $%
H_{c}^{l=1.1}(\tau =10^{-30}s)=2.40\times 10^{35}s^{-1}$. Taking into
account (\ref{tau}) one obtains that the condition (\ref{eqH1taul=1.1}) is
satisfied for $\tau ^{-1}<H_{1}<H_{c}^{l=1.1}(\tau )$.
\end{enumerate}

\bigskip

(ii) When $l=2$ (expansion dominated by the MBHs) we have 
\begin{equation}
f(2,H_{1},\tau )=-51.89+4\log _{10}\left( \frac{H_{1}}{1s^{-1}}\right)
-2\log _{10}\left( 1+\frac{l+1}{l}H_{1}\tau \right) <0,  \label{eqH1taul=2}
\end{equation}%
see figure \ref{graphf}. Inspection of (\ref{eqH1taul=2}) and Fig. \ref%
{graphf} reveals that:

\begin{enumerate}
\item For $\tau <\tau _{c}^{l=2}=6.75\times 10^{-14}s$, one obtains $%
H_{1}<10^{13}s^{-1}$ which is totally incompatible with condition (\ref{tau}%
), $H_{1}>1.48\times 10^{13}s^{-1}$ for $\tau =\tau _{c}^{l=2}$ in the most
favorable case. Thus, the region $\tau <\tau _{c}^{l=2}$ is ruled out as
predicts an excess of anisotropy in the CMBR.

\item For $\tau >\tau _{c}^{l=2}$, one obtains $H_{1}<H_{c}^{l=2}(\tau )$ from the 
condition (\ref{eqH1taul=2}). $H_{c}^{l=2}(\tau )$ is
always larger than $1.48\times 10^{13}s^{-1}$ for $\tau $ in the range
considered, e.g. $H_{c}^{l=2}(\tau =10^{-10}s)=1.35\times 10^{16}s^{-1}$.
Conditions (\ref{tau}) and (\ref{eqH1taul=2}) are both satisfied in this
range for $\tau ^{-1}<H_{1}<H_{c}^{l=2}(\tau )$.
\end{enumerate}

\bigskip

We may conclude that the condition (\ref{eqH1taul}) leads to different
allowed ranges for $H_{1}$ and $\tau $ for each $l$ considered, although
their interpretation is rather similar. For $\tau <$ $\tau _{c}^{l}$ the
condition of minimum duration of the \ `MBHs+rad' era (\ref{tau}) and the
upper bound given by the CMBR anisotropy are incompatibles. However, for $\tau >$ $%
\tau _{c}^{l}$ these two conditions are compatible for $\tau
^{-1}<H_{1}<H_{c}^{l}(\tau )$. 

\section{Concluding remarks}

We have calculated the power spectrum of RGWs in a universe
that begins with an inflationary phase, followed by a phase dominated
by a mixture of MBHs and radiation, then a radiation dominated phase (after
the MBHs evaporated), and finally a dust dominated phase. The spectrum depends 
just on three free parameters, namely $H_{1}$ the Hubble factor at the transition inflation-`MBHs+rad' 
era, $\tau $, the cosmological time span of the `MBHs+rad' era, and the power $l$, 
being $a(\eta )\propto\eta ^{l}$ the scale factor of the `MBHs+rad' era with $1<l\leq 2$.

The upper bound on the spectrum of RGWs obtained from the CMBR anisotropy places
severe constraints on $H_{1}$ and $\tau $. For each value of $l$
considered, there is a minimum value of $\tau $, $\tau _{c}^{l}$, compatible
with the CMBR anisotropy. There is a range of $\tau $, $\tau >\tau _{c}^{l}$%
, for which $\tau ^{-1}<H_{1}<H_{c}^{l}(\tau )$ satisfies the CMBR
upper bound.

The four-stage scenario predicts a much lower power spectrum of RGWs than
the conventional three-stage scenario. We may therefore conclude that if
LISA fails to detect a spectrum at the level expected by the three stage model,
rather than signaling than the recycle model of Khoury \textit{et al.} \cite%
{Khoury}\ should supersede the standard big-bang inflationary model it may
indicate a MBHs+radiation era between inflation and radiation dominance truly 
took place. Likewise, once the spectrum is successfully measured we
will be able to learn from it the proportion of MBHs and radiation in the
mixture phase.

In reality the current epoch is not dust dominated as the recent supernovae
type Ia data seems to suggest that the Universe is dominated by a dark
energy component and cold dark matter implying an epoch of accelerated
expansion \cite{Perl et al}. This implies a new transition in the scale
factor shape at some time $\eta _{4}$ with $\eta _{3}<\eta _{4}<\eta _{0}$.
The spectrum of RGWs with $k_{0}<k<k_{4}$ is changed. Thus, the constraints
on $H_{1}$ and $\tau $ obtained from the CMBR bound stay unchanged as the
RGWs created at the transition dust-accelerated era were not present at the
last scattering.

\acknowledgments
GI was supported by a ``Beca Predoctoral de Formaci\'{o} d'Investigadors de la UAB"
(Convocatoria any 2001). This work was partially supported by the Spanish Ministry 
of Science and Technology under grants BFM 2000-C-03-01 and 2000-1322.

\end{document}